# SYMBOLIC MUSIC LOOP GENERATION WITH NEURAL DISCRETE REPRESENTATIONS

**Sangjun Han[1], Hyeongrae Ihm[1], Moontae Lee[1,2], Woohyung Lim[1]**
[1] LG AI Research, [2] University of Illinois at Chicago
{sj.han, hrim, moontae.lee, w.lim}@lgresearch.ai

## ABSTRACT

Since most of music has repetitive structures from motifs to phrases, repeating musical ideas can be a basic operation for music composition. The basic block that we focus on is conceptualized as loops which are essential ingredients of music. Furthermore, meaningful note patterns can be formed in a finite space, so it is sufficient to represent them with combinations of discrete symbols as done in other domains. In this work, we propose symbolic music loop generation via learning discrete representations. We first extract loops from MIDI datasets using a loop detector and then learn an autoregressive model trained by discrete latent codes of the extracted loops. We show that our model outperforms well-known music generative models in terms of both fidelity and diversity, evaluating on random space. Our code and supplementary materials are available at https://github.com/sjhan91/Loop_VQVAE_Official.

## 1. INTRODUCTION

With the advance of generative models, many studies are trying to model sequential data such as language and speech. Music can also benefit from their previous works since it consists of a sequence of multiple notes to represent the composer's intention. Through the advancement, individuals can imitate the musical inspiration of artists without musical expertise.

Several works related to music generation have focused on generating long sequences by utilizing the expressive power of Transformer [1-4]. It is a promising approach because the Transformer with hierarchical layers can learn various types of repetitive structures on its self-attentions. However, it still has limitations, derived from the error accumulation and rhythmic irregularity, to achieve the ultimate goal which is to compose full-length music [1, 2]. To tackle that problem, we explicitly utilize recurrence properties in music, generating short and fixed-length music phrases that can be used as basic patterns of music.

Repeating musical ideas is a basic operation for music composition. This operation conceptualizes the loop, an essential ingredient for creating remixes or mash-ups [5]. With the concept, we can simplify the generation task into generating one distinctive pattern which consists only of a few bars. For loop extraction, previous works have attempted to detect autocorrelated peaks (it relies on an assumption that it is sufficient to detect the starting point of phrases) [5, 6], but we apply the knowledge of overall loop structures, obtained from a public audio dataset, to the MIDI domain.

Another intuition is that the musical ideas can be formed in combinations of finite symbols. It is known that learning discrete latent codes is sufficient to represent the continuous world since many modalities consist of sequences of symbols [7]. For example, objects in vision, words in language, and phonemes in speech may be candidates of symbols. Also, compressing raw data into discrete semantic units makes an autoregressive model easy to train by capturing long-range data dependency [8]. If we regard consecutive notes as symbols, it is natural to adopt discrete representation as the basis of our autoregressive generator. This process can emphasize pattern to pattern structures sacrificing the minimal loss of note details.

In contrast to using Inception Score (IS) and Fréchet Inception Distance (FID) for computer vision [9, 10], no consensus has been made for evaluating generated music. This is because the absence of pre-trained feature extractors prohibits the comparison of true and generated samples on feature space. Recently, Naeem has shown that random embedding can also be effective when the target distribution is far from pre-trained model statistics [11]. Using the random initialized networks, we evaluate our generative models on sample fidelity and diversity as firstly suggested in [12].

In this work, we propose symbolic music loop generation via learning discrete representations. It involves two main processes; 1) *loop extraction* from MIDI datasets using a loop detector and 2) *loop generation* from an autoregressive model trained by discrete latent codes of the extracted loops. The outputs of the generative model are loops consisting of 8 bars, which can be repeated seamlessly. Since we aim to generate polyphony and multitrack sequences, the bass and drum are chosen for our experiments, which are fundamental components of melody and rhythm. Additionally, we adopt an evaluation protocol from [11] to measure two-dimensional score which stands for fidelity and diversity. Our contributions are summarized as follows;

- We propose the framework of symbolic music loop generation, which involves loop extraction, loop generation, and its evaluations.

- For loop extraction, we design a structure-aware loop detector trained by external audio sources to extract loops of 8 bars from MIDI.
- For loop generation, we verify that an autoregressive model combined with discrete representations can generate plausible loop phrases which can be repeated.
- With randomly initialized networks for embedding, we evaluate sample quality in terms of fidelity and diversity.

## 2. RELATED WORK

We introduce several works related to the history of loop extraction and music generation, the effectiveness of discrete representations, and the development of evaluating generative models.

### 2.1 Loop Extraction

For symbolic music generation, some researchers have prepared their dataset by sliding a window with a stride of 1 bar [13, 14]. Although they have achieved good generative performance, this process does not consider relative positions within music, generating ambiguous phrases.

Some works have imposed structural constraints on music generation models [15, 16], or directly detect novel segments which are repetitive in time series [17, 18]. In the audio domain, there have been attempts to extract loops explicitly by capturing repeated phrases [5, 6]. They extract harmonic features such as chroma vectors or mel-frequency cepstrum and catch autocorrelation peaks to determine the starting point of loops. They also require a heuristic process to decide which features should be more weighted. In contrast, our loop detector works on a data-driven approach, so it does not require a manual process such as feature extraction and weighting strategies.

A recent work for drum loop generation has informed the availability of a human-created loop dataset from Looperman (https://www.looperman.com/) [19]. Although it consists of audio sources with various instruments and genres, we suggest a promising approach to combine them with Lakh MIDI Dataset (LMD) [20]. Concretely, we extract domain-invariant loop structures from Looperman and train a loop detector using them to extract loops from LMD.

### 2.2 Symbolic Music Generation

To make MIDI available in machine learning, two representation methods are prevalent; event-based representation and time-grid based representation [21]. Although the former can represent high time-resolution with a few event vectors, we choose the latter one to benefit from fixed-length and repetitive structures for music. There have been several works to deal with polyphony multitrack representation [1, 13, 22]. Especially for time-grid based representation, MuseGAN [22] has stacked five instruments each of which consists of 4 bars and 84 pitches. We follow their method while restricting to two instruments, the bass and drum.

### 2.3 Discrete Representations

As opposed to VAE [23] with continuous prior, Oord has proved that a finite set of latent codes is sufficient to reconstruct while it prevents posterior collapse [7]. Powerful autoregressive priors with the latent codes have shown promising performance not only in image generation [8] but also text2image [24], and video generation [25]. Loop generation can also benefit from discrete data compression by expressing long-range structural patterns.

### 2.4 Evaluation of Generative Models

Proper evaluation metrics for generative models are important to reduce human evaluation labor. Previous works of music generation have inspected model metrics (*e.g.,* log-likelihood) or musical metrics on data space to evaluate how much true and generated samples are similar [3, 22, 26]. Comparison on data space, however, is vulnerable to pixel by pixel phase difference, ignoring data semantics. Without available pre-trained networks, it has been reported that random embeddings are more robust for evaluation rather than using models trained by other data domains [11]. In this respect, we take randomly initialized networks as our feature extractor and evaluate our music samples on feature space.

The commonly used metrics in computer vision are IS and FID which compute a one-dimensional score. To distinguish fidelity and diversity from the score, Sajjadi has suggested precision and recall evaluating overlapped ratio between true and generated distribution [12]. After that, some works have proposed different ways of constructing data distribution using k-nearest neighbors (KNN) [11, 27]. We adopt KNN based evaluation protocol since it is not affected by its initialization and is robust to outliers.

## 3. PROPOSED METHOD

Our work starts with collecting MIDI loops using a structure-aware loop detector. Since we design the detector to take not the music itself but bar-to-bar structures, we can train it with a loop-labeled dataset from audio domain. After obtaining the MIDI loop set, we design a loop generator in two-stage; compressing data into discrete space and building an autoregressive model with them. Lastly, generated samples are evaluated on qualitative and quantitative metrics.

### 3.1 Data Preparation

*3.1.1 Looperman Dataset*

Looperman Dataset from the audio domain is collected and transformed for training our loop detector. We collect 1,000 loops of 8 bars from Looperman, a website allowing to upload and download free music loops. Formally, we denote one loop that is 1-D audio as $x_{WAV} \in \mathcal{R}$. The process of data transformation for the loop detector will be described in section 3.2.1.

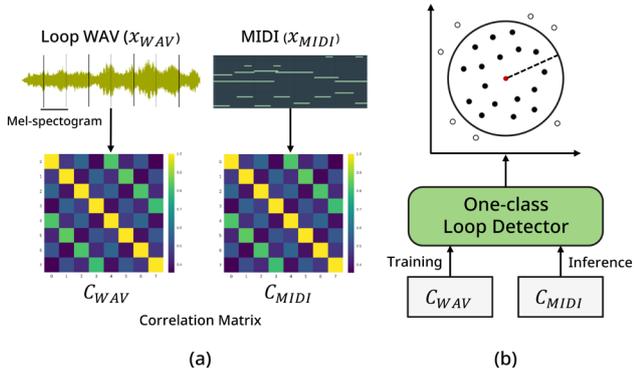
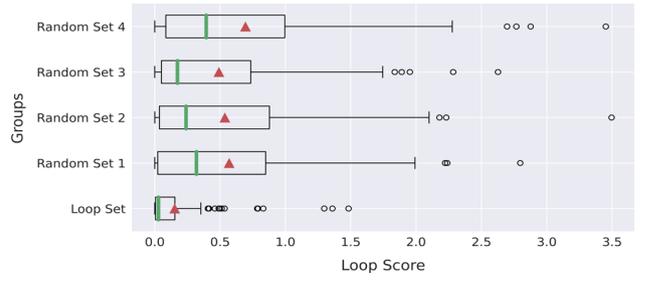

**Figure 1.** The process of loop extraction. (a) $x_{WAV}$ and $x_{MIDI}$ are transformed to each correlation matrix. (b) $C_{WAV}$ are used to train the one-class loop detector and we extract loops from Lakh MIDI Dataset by forward passing $C_{MIDI}$ to the detector.

**Figure 2.** The evaluation of our loop detector. Green lines in the boxes indicate median values and red triangles for mean values.

*3.1.2 Lakh MIDI Dataset*

Lakh MIDI Dataset is a collection of MIDI files with various genres and tracks, so it is appropriate to conduct symbolic music experiments [20]. In this experiment, each note is quantized on the 16th note unit with binary representation so that 16 notes are placed in a bar. To verify the feasibility of multitrack polyphony generation, we extract two instruments; bass guitar (program=32~39) and drum (is_drum=True) which play crucial roles of melodic and rhythmic patterns in music. The bass pitches are clipped from C1 to B4 (48 pitches). If several pitches for the bass are played at the same time, we make only the lowest pitch alive for natural play. For the drum set, nine components (kick, snare, closed hi-hat, open hi-hat, low tom, mid tom, high tom, crash, and ride) are regarded as a standard set and the rest of the components are incorporated into the closest one or discarded. Formally, our pianoroll representation can be described as follows; $x_{MIDI} \in \{0,1\}^{(T \times B) \times P}$ where $T$ is the number of time steps in a bar ($T = 16$), $B$ is the number of bars ($B = 8$), and $P$ is the number of pitches ($P = 57$). We collect 5,687,274 phrases of 8 bars by sliding a window with a stride of 1 bar, removing non-4/4 signature music. Using pretty_midi [28] and pypianoroll [29] in Python library, MIDI processing is conducted.

### 3.2 Loop Extraction

*3.2.1 Data Transformation for the Loop Detector*

We transform each the $x_{WAV}$ and $x_{MIDI}$ to $B \times B$ matrix indicating bar-to-bar correlation (Figure 1 (a)). For the $x_{WAV}$, we extract $B$ mel-spectrograms each corresponding to $B$ bars and compute a correlation matrix ($C_{WAV}$). For the $x_{MIDI}$, we compute normalized Hamming distance among bars and renormalize it to express correlation ($C_{MIDI}$). Only the upper triangle part of $C$ is used. More details are described in Appendix B.3.

*3.2.2 Loop Extraction through the Loop Detector*

Our loop detector is to classify loop and non-loop phrases, given only loop-labeled datasets. It is related to the problem of anomaly detection trained in an unsupervised way to construct normal data distribution. At inference time, outliers from the distribution are regarded as anomalous samples. Similarly, we treat $C_{WAV}$ as normal samples (training set) and measure the likelihood of $C_{MIDI}$ (test set) on $C_{WAV}$ distribution. Among several ways, we choose One-Class Deep SVDD [30] as our loop detector, of which the training objective is

$$min \; \frac{1}{n}\sum_{i=1}^{n} ||f_w(x_i) - c||^2 + \lambda \Omega(W) \qquad (1)$$

where $f_w$ is a neural network taking input $x$ with learnable parameters $W$, $n$ is the number of training samples, $c$ is a center vector, and $\Omega(W)$ is a controllable regularizer. The objective can be thought as mapping all data samples close to center $c$, contracting a hypersphere. Initially, $f_w$ is trained to reconstruct $x$ with a decoder $f_w^{-1}$ and center $c$ is set to mean vectors acquired from $f_w$ initial pass of the training data. $f_w$ consists of 3 fully-connected layers with bias-off and LeakyReLU(0.1) to prevent hypersphere collapse as referred in [30]. During training, AdamW optimizer [31] is applied with cosine annealing from 1e-3 to 5e-6 for 1,000 epochs. At inference time, the loop score of $C_{MIDI}$ can be obtained as

$$loop \; score = ||f_{w^*}(x) - c||^2 \qquad (2)$$

where $w^*$ stands for optimized parameters of $f$ (the process of loop extraction is illustrated on Figure 1 (b)). Samples with lower loop scores are considered close to the loop.

To evaluate the loop detector, we manually pick 100 loop samples from $x_{MIDI}$ and compare them with 4 groups randomly picked (Figure 2). Although the randomized groups can contain subsets of loops, we can verify that the loop set group indicates the lowest loop score (0.153 ($\pm$0.282)). Also, paired *t*-test between the loop group and each other group shows the statistically significant difference with $p < 0.001$. When determining whether loop or not, we set a conservative threshold as positive one sigma of the loop score distribution (transformed by $log$ to fit close to Gaussian distribution) from the training set. Consequently, we collect 751,935 $x_{MIDI}$ to be used at loop generation stage.

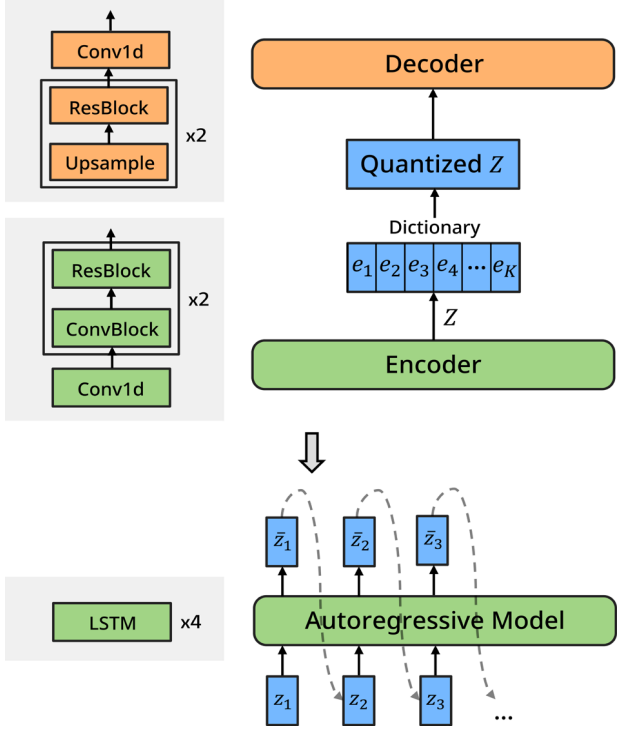

**Figure 3.** The process of loop generation. The upper denotes VQ-VAE and the bottom for LSTM based autoregressive model.

### 3.3 Loop Generation

*3.3.1 Data Compression using VQ-VAE*

VQ-VAE [7] maps a data sequence into discrete latent space and reconstructs it to the original data space. With an encoder $q_\phi$ and a decoder $p_\theta$, the objective becomes

$$max\ \mathbb{E}[log\ p_\theta(x|z)] - \beta||q_\phi(z|x) - sg[e]||\quad (3)$$

where $sg$ denotes a stop gradients operator for the dictionary embedding $e$. During forward pass, latent $z$ from the encoder $q_\phi(z|x)$ is quantized to nearest embedding $e$. The second term above is responsible for the latent $z$ not to diverge far from $e$. For every batch, the embedding dictionary is updated in the sense of centroids of K-means clustering.

Our VQ-VAE (Figure 3) encodes $x = \{x_1, x_2, \cdots, x_{T \times B}\}$, $x_i \in \mathbb{R}^P$ into $z = \{z_1, z_2, \cdots, z_S\}$, $z_i \in \mathbb{R}^D$ where S denotes the number of time steps ($S = 32$) in latent space and $D$ denotes latent dimensions ($D = 16$). Starting from randomly initialized dictionary $e = \{e_1, e_2, \cdots, e_K\}$, $e_i \in \mathbb{R}^D$ ($K = 512$), the latent $z_i$ is mapped to the nearest embedding $e_k$ where $k = argmin_j ||z_i - e_j||$. After that, the embeddings are passed to the VQ-VAE decoder to reconstruct their original data. For the reconstruction objective, our data representation is regarded as multi-label for each time step (multiple 1s can exist on pitch $P$ dimension at one time step), so cross-entropy loss with softmax is not suitable. Our objective for the reconstruction is described as

$$min\ -\frac{1}{nm}\sum_{i=1}^{n}\sum_{j=1}^{m} L\left(\sigma(o_{ij}), y_{ij}\right)\quad (4)$$

where $L$ denotes binary cross-entropy, $\sigma$ is sigmoid function, $o$ is outputs of the VQ-VAE with $m$ notes, and $y$ for ground truth. When $\sigma(o_{ij}) \geq 0.5$, it predicts label as 1, otherwise 0.

| Model | Reconstruction Error |
|---|---|
| CNN-VAE | 4.412e-3 |
| VQ-VAE | 6.643e-3 |

**Table 1.** The reconstruction errors. This is computed as the average of hamming distance between input and target samples of the validation set.

The details of our VQ-VAE architectures (Figure 3) are described in Appendix B.4. Empirically, we have verified that a high compression ratio, especially along $S$ dimension, severely deteriorates the reconstruction task of the VQ-VAE. The value of $S$ has been determined to consider both the compression ratio and reconstruction quality. AdamW optimizer and cosine annealing from 1e-3 to 5e-6 is applied to the model training. As did in [32], random restarting is applied to less referenced $e_i$, replacing them with one of batch samples. After training, we forward pass the training set to obtain quantized embeddings $z$.

*3.3.2 Generation through an Autoregressive Model*

We design an autoregressive model $\prod_{i=0}^{S} p_\theta(z_i|z_{<i})$ over the quantized embeddings to generate unseen samples. The quantized indices $k$ are used as inputs of 4-layers LSTM with an embedding layer. For each time step, softmax output predicts next step index $k$ (validation accuracy 76.651% in teacher forcing mode). $z_0$ is sampled from multinomial distribution $p(z_0)$ of the training set. When sampling unseen data in full sampling mode, temperatures in softmax enable us to control sample diversity (temperature=0.7). We compare several sampling methods such as temperature sampling, top-k sampling [33], and nucleus sampling [34]. The generated indices are decoded by the VQ-VAE's decoder.

### 4. QUANTITATIVE EVALUATION

For quantitative evaluation of generated samples, we address three concepts; 1) *model metric* related to evaluating the capacity of generative models, 2) *musical style* related to measuring how much our intended properties of the loop are involved in generated samples, and 3) *similarity metrics* related to how much generated samples are involved in the training set on feature space (or vice versa).

### 4.1 Model Metric

**Reconstruction Error:** For VAE, the reconstruction error is part of the objective function that indicates how well the model decodes its latent features to the original data. However, perfect satisfaction with the objective does not guarantee to generate high-quality samples (empirically, we have verified that original VAE has produced many noisy samples even after achieving the minimal reconstruction error when forcing Kullback-Leibler divergence term to 0).

| Model | LS | UP | ND | P | R | D | C |
|---|---|---|---|---|---|---|---|
| Training Set | 6.806e-3 | 5.769 | 14.383 | - | - | - | - |
| CNN-VAE | 3.496e-1 | 5.614 | 12.648 | 0.642±0.033 | 0.625±0.021 | 0.617±0.076 | 0.746±0.024 |
| Music Transformer | 7.200e-1 | 4.127 | 11.290 | 0.546±0.077 | 0.359±0.113 | 0.687±0.174 | 0.408±0.064 |
| MuseGAN | 2.307e-1 | **5.790** | 14.011 | 0.641±0.013 | **0.689±0.012** | 0.673±0.045 | 0.842±0.013 |
| VQ-VAE+LSTM (temperature sampling) | 2.275e-1 | 5.079 | 14.289 | 0.768±0.013 | 0.655±0.022 | 1.263±0.047 | 0.949±0.002 |
| VQ-VAE+LSTM (top-k sampling=30) | **1.978e-1** | 5.044 | 14.320 | 0.779±0.015 | 0.636±0.015 | 1.328±0.072 | **0.952±0.004** |
| VQ-VAE+LSTM (top-p sampling=0.08) | 2.037e-1 | 5.042 | **14.341** | **0.783±0.017** | 0.638±0.029 | **1.337±0.075** | 0.950±0.005 |

**Table 2.** The result table indicates all metrics explained at section 4.2 and 4.3. We compute P, R, D, and C from 10 different networks, so we denote mean values with standard deviations. Best values are marked in bold font.

| Model | $F_1$ score (P & R) | $F_1$ score (D & C) |
|---|---|---|
| CNN-VAE | 0.633 | 0.675 |
| Music Transformer | 0.433 | 0.512 |
| MuseGAN | 0.664 | 0.748 |
| VQ-VAE+LSTM (temperature sampling) | **0.707** | 1.084 |
| VQ-VAE+LSTM (top-k sampling=30) | 0.700 | 1.109 |
| VQ-VAE+LSTM (top-p sampling=0.08) | 0.703 | **1.111** |

**Table 3.** F1 score from Table 2 results.

### 4.2 Musical Style

The investigation of used harmonics and rhythm patterns indicates the musical style of our generated samples.
**Loop Score (LS):** Using the trained loop detector, we can evaluate how much generated samples are close to the loop.
**Unique Pitch (UP):** It computes the average number of used pitches per bar [22]. It reflects harmonic components on data space. It is desirable for generated samples to follow UP values of the training set.
**Note Density (ND):** It computes the average number of notes played per bar considering all instruments [26]. It reflects rhythmic components on data space. It is desirable for generated samples to follow ND values of the training set.

### 4.3 Similarity Metrics

Evaluating generated music samples on data space considers only musical features that humans can perceive. Here, we measure similarity of true and generated samples on latent space while indicating the sample fidelity and diversity.
**Precision & Recall (P & R):** The fidelity of generative models can be evaluated on precision which measures how much generated distributions are involved in true distributions. Likewise, the diversity can be realized by recall which measures how much true distributions are involved in generated distributions [12].

Kynkäänniemi have proposed improved P & R that count the presence of samples on the overlapped data manifold constructed by KNN [27]. In the case of precision, the data manifold is constructed from multiple spheres whose center is determined by true samples and whose radius is the distance between the true samples and their k-th nearest neighbors. All operations in the P & R should be conducted on latent space, so we use simple CNN networks initialized randomly for embedding [11]. For both P & R, we fix $k$ of the KNN to 5 and get the average of the metrics from 10 different networks. To avoid extensive computation of the KNN, we use 10,000 samples for each network.
**Density and Coverage (D & C):** P & R are vulnerable to outliers overestimating data manifold. To remedy this, [11] have counted the average number of overlapped samples on a sphere. The concept and process of D & C are similar with P & R, except that the density can be greater than 1.

## 5. EXPERIMENTS

We evaluate our model by comparing it to 1) *training set*, 2) *CNN-VAE* similar structure with our VQ-VAE, 3) *Music Transformer* [3], and 4) *MuseGAN* [22]. All models have generated loop samples as many as the training set. Additionally, we apply several sampling methods for the VQ-VAE+LSTM and compare them. The experiment details of the baselines are explained in Appendix A.

### 5.1 Quantitative Evaluation Results

*5.1.1 Model Metric Results*

Due to the finite latent codes, our VQ-VAE is a little worse than the CNN-VAE for the reconstruction task (Table 1).

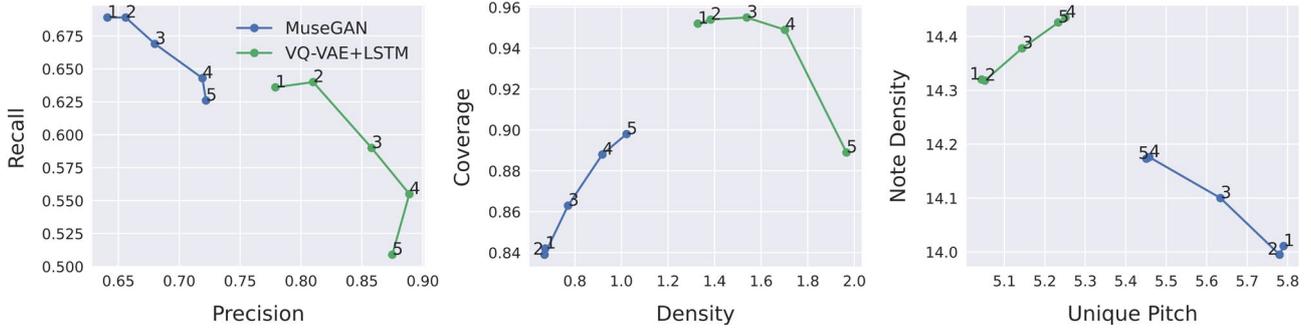

**Figure 4.** Rejection sampling results for the MuseGAN and VQ-VAE+LSTM. Each numbering markers indicates the various rejection rates (getting stricter from 1 to 5). The rates correspond to {no apply, 1, 0.1, 0.01, 0.001}.

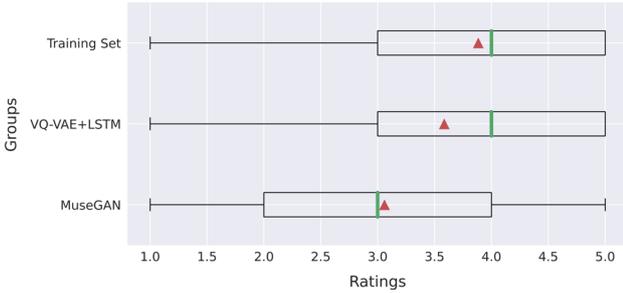

**Figure 5.** Human listening test. Green lines in the boxes indicate median values and red triangles for mean values.

As referred in [7], we have observed that $\beta$ in the VQ-VAE objectives did not affect much to the task ($\beta = 0.25$).

*5.1.2 Musical Metric Results*

Our proposed model achieves the highest performance in terms of LS and ND (Table 2). The reason for poor performance in UP may be related to the VQ-VAE's reconstruction performance. Nevertheless, they can generate highly structural and rhythmic samples in a holistic view. Note that the Music Transformer fails to preserve the loop properties in its generated samples.

*5.1.3 Similarity Metric Results*

Except for the recall, our model achieves much better scores in terms of precision, density and coverage (Table 2). In MuseGAN, the recall may be overestimated by generated outliers. Depending on sampling methods and their parameters for full sampling mode, we can observe that there is a trade-off between fidelity and diversity. Even with the random embeddings, the all metrics show consistent performance (low standard deviations). A comprehensive evaluation ($F_1$ score) can be found in Table 3.

### 5.2 Rejection Sampling

It is promising that the loop detector can be used to control the trade-off between fidelity and diversity of generative models. This concept, rejection sampling, is to reject generated samples which do not meet our conditions. (loop scores in this context) [8]. If setting the rejection rate strictly, we can obtain music samples which are closer to the loop. For the experiments, we choose high-scored models that are MuseGAN and VQ-VAE+LSTM with top-k sampling. Figure 4 shows the P & R, D & C, and UP & ND results for various rejection rates. In P & R, the two models show similar trends while applying more strict rejection rates, but opposite trends for other metrics. Contrary to our assumption, both D & C of MuseGAN increase as we apply the stricter loop detector. It rather disproves that MuseGAN produces many outliers, so the loop detector may help to increase sample diversity close to the true distribution. In terms of ND, both models indicate minimal effect with the rejection sampling. However, they show contradiction about the aspect of UP changes.

### 5.3 Human Listening Test

We conduct a listening test for 20 people. We select the training set as a baseline and compare loop samples from two generative models (MuseGAN and VQ-VAE+LSTM with top-k sampling). Participants are asked to listen 10 samples for each group (total 30 samples) and evaluate how much the sample sounds like the loop music (which can be repeated seamlessly) on a Likert scale.

As Figure 5, the training set group achieves the highest ratings with an average rating of 3.885 ($\pm 1.045$). For the generative models, VQ-VAE+LSTM achieves the highest average rating 3.585 ($\pm 1.141$). It seems that the participants have felt them closer to the real music since loops from discrete representations are repetitive and structural (MuseGAN 3.060 ($\pm 1.172$)). Additionally, we carry out Kruskal-Wallis H test which is non-parametric one-way ANOVA. The test shows a statistically significant difference among the test groups with $H = 49.811$, $p < 0.001$.

### 6. CONCLUSION

We leverage recurring nature of music by adopting the concept of the loop. To fulfill our objective, we address two processes; *loop extraction* and *loop generation*. First, we design a loop detector trained by a loop audio dataset to prepare loops from MIDI. Second, we adopt the two-stage generative approach, compressing data into discrete representations and designing an autoregressive model. Even without pre-trained feature extractors, we can evaluate our generative models on measuring fidelity and diversity. It is observed that our model outperforms well-known generative model for loop generation.